\documentclass[aps,prl,superscriptaddress,showpacs,floatfix,twocolumn]{revtex4}

\usepackage{amssymb, amsmath}
\usepackage[dvips]{graphicx}
\def\v2{\mbox{$v_2$}}

\def\sqrtsNN{\mbox{$\sqrt{s_{NN}}$}}
\def\agt{\underset{\raise0.3em\hbox{$\smash{\scriptscriptstyle\thicksim}$}}{ > }}
\def\alt{\underset{\raise0.3em\hbox{$\smash{\scriptscriptstyle\thicksim}$}}{ > }}

\begin{document}


%
\hyphenation{author another created financial paper re-commend-ed Post-Script}

\title{Three-dimensional two-pion source image from Pb+Pb collisions at \sqrtsNN=17.3~GeV: 
                 new constraints for source breakup dynamics}

%
%

\author{
C.~Alt$^{9}$, T.~Anticic$^{23}$, B.~Baatar$^{8}$,D.~Barna$^{4}$,
J.~Bartke$^{6}$, L.~Betev$^{10}$, H.~Bia{\l}\-kowska$^{20}$,
C.~Blume$^{9}$,  B.~Boimska$^{20}$, M.~Botje$^{1}$,
J.~Bracinik$^{3}$, P.~Bun\'{c}i\'{c}$^{10}$,
V.~Cerny$^{3}$, P.~Christakoglou$^{1}$,
P.~Chung$^{19}$, O.~Chvala$^{14}$,
J.G.~Cramer$^{16}$, P.~Csat\'{o}$^{4}$, P.~Dinkelaker$^{9}$,
V.~Eckardt$^{13}$,
D.~Flierl$^{9}$, Z.~Fodor$^{4}$, P.~Foka$^{7}$,
V.~Friese$^{7}$, J.~G\'{a}l$^{4}$,
M.~Ga\'zdzicki$^{9,11}$, V.~Genchev$^{18}$,
E.~G{\l}adysz$^{6}$, K.~Grebieszkow$^{22}$,
S.~Hegyi$^{4}$, C.~H\"{o}hne$^{7}$,
K.~Kadija$^{23}$, A.~Karev$^{13}$,
S.~Kniege$^{9}$, V.I.~Kolesnikov$^{8}$,
R.~Korus$^{11}$, M.~Kowalski$^{6}$,
M.~Kreps$^{3}$, A.~Laszlo$^{4}$,
R.~Lacey$^{19}$, M.~van~Leeuwen$^{1}$,
P.~L\'{e}vai$^{4}$, L.~Litov$^{17}$, B.~Lungwitz$^{9}$,
M.~Makariev$^{17}$, A.I.~Malakhov$^{8}$,
M.~Mateev$^{17}$, G.L.~Melkumov$^{8}$, M.~Mitrovski$^{9}$,
J.~Moln\'{a}r$^{4}$, St.~Mr\'owczy\'nski$^{11}$, V.~Nicolic$^{23}$,
G.~P\'{a}lla$^{4}$, A.D.~Panagiotou$^{2}$, D.~Panayotov$^{17}$,
A.~Petridis$^{2,\dagger}$, W.~Peryt$^{22}$, M.~Pikna$^{3}$, J.~Pluta$^{22}$, D.~Prindle$^{16}$,
F.~P\"{u}hlhofer$^{12}$, R.~Renfordt$^{9}$,
C.~Roland$^{5}$, G.~Roland$^{5}$,
M. Rybczy\'nski$^{11}$, A.~Rybicki$^{6}$,
A.~Sandoval$^{7}$, N.~Schmitz$^{13}$, T.~Schuster$^{9}$, P.~Seyboth$^{13}$,
F.~Sikl\'{e}r$^{4}$, B.~Sitar$^{3}$, E.~Skrzypczak$^{21}$, M.~Slodkowski$^{22}$,
G.~Stefanek$^{11}$, R.~Stock$^{9}$, H.~Str\"{o}bele$^{9}$, T.~Susa$^{23}$,
I.~Szentp\'{e}tery$^{4}$, J.~Sziklai$^{4}$, M.~Szuba$^{22}$, P.~Szymanski$^{10,20}$,
V.~Trubnikov$^{20}$, D.~Varga$^{4,10}$, M.~Vassiliou$^{2}$,
G.I.~Veres$^{4,5}$, G.~Vesztergombi$^{4}$,
D.~Vrani\'{c}$^{7}$, Z.~W{\l}odarczyk$^{11}$, A.~Wojtaszek$^{11}$, I.K.~Yoo$^{15}$
                        \\  ( NA49 Collaboration ) \\
J.M.~Alexander$^{19}$, P.~Danielewicz$^{24,25}$, A. Kisiel$^{22,26}$, S.~Pratt$^{25}$
}

\affiliation{
$^{1}$NIKHEF, Amsterdam, Netherlands. \\
$^{2}$Department of Physics, University of Athens, Athens, Greece.\\
$^{3}$Comenius University, Bratislava, Slovakia.\\
$^{4}$KFKI Research Institute for Particle and Nuclear Physics, Budapest, Hungary.\\
$^{5}$MIT, Cambridge, USA.\\
$^{6}$Henryk Niewodniczanski Institute of Nuclear Physics, Polish Academy of Sciences, Cracow, Poland.\\
$^{7}$Gesellschaft f\"{u}r Schwerionenforschung (GSI), Darmstadt, Germany.\\
$^{8}$Joint Institute for Nuclear Research, Dubna, Russia.\\
$^{9}$Fachbereich Physik der Universit\"{a}t, Frankfurt, Germany.\\
$^{10}$CERN, Geneva, Switzerland.\\
$^{11}$Institute of Physics \'Swi\c{e}tokrzyska Academy, Kielce, Poland.\\
$^{12}$Fachbereich Physik der Universit\"{a}t, Marburg, Germany.\\
$^{13}$Max-Planck-Institut f\"{u}r Physik, Munich, Germany.\\
$^{14}$Charles University, Faculty of Mathematics and Physics, Institute of Particle and Nuclear Physics, Prague, Czech Republic.\\
$^{15}$Department of Physics, Pusan National University, Pusan, Republic of Korea.\\
$^{16}$Nuclear Physics Laboratory, University of Washington, Seattle, WA, USA.\\
$^{17}$Atomic Physics Department, Sofia University St. Kliment Ohridski, Sofia, Bulgaria.\\
$^{18}$Institute for Nuclear Research and Nuclear Energy, Sofia, Bulgaria.\\
$^{19}$Department of Chemistry, SUNY Stony Brook, Stony Brook, NY, USA.\\
$^{20}$Institute for Nuclear Studies, Warsaw, Poland.\\
$^{21}$Institute for Experimental Physics, University of Warsaw, Warsaw, Poland.\\
$^{22}$Faculty of Physics, Warsaw University of Technology, Warsaw, Poland.\\
$^{23}$Rudjer Boskovic Institute, Zagreb, Croatia.\\
$^{24}$National Superconducting Cyclotron Laboratory, MSU, East Lansing, MI, USA.\\
$^{25}$Department of Physics and Astronomy, Michigan State University,
    East Lansing, MI, USA.\\
$^{26}$Department of Physics, The Ohio State University, Columbus, Ohio 43210.\\
 }
%
%
%
\date{\today}
\begin{abstract}

Source imaging methodology is used to provide a three-dimensional two-pion source function for
mid-rapidity pion pairs with $p_T<70$~MeV/c in central ($0-7\%$) Pb+Pb collisions 
at $\sqrt s_{NN}$=17.3~GeV. Prominent non-Gaussian tails are observed in the pion pair transverse 
momentum (outward) and in the beam (longitudinal) directions. Model calculations 
reproduce them with the assumption of Bjorken longitudinal boost invariance and transverse flow 
blast-wave dynamics; 
they also yield a proper time for breakup and emission duration for the pion source.
%

%
\end{abstract}

%
\maketitle

Over the last several decades, the expansion dynamics and breakup lifetime for the 
exotic matter produced in relativistic heavy ion (RHI) collisions, have been of paramount 
interest \cite{shu05,pra84}. Such enormous energy densities are created in the RHI 
collision zone, that deconfinement of nuclear matter is expected \cite{adc05}. 
To gain an understanding of this state of matter, 
it is essential to study its dynamical evolution. The space-time extent of hot particle 
emission sources in heavy ion collisions has been studied for years via final-state 
correlations between two particles \cite{lis05}.

Years ago, pioneering work at the Alternating Gradient Synchrotron (AGS) \cite{Chapman:1996ec} 
and by the NA49 Collaboration at the CERN Super Proton Synchrotron (SPS)  \cite{app98}, exploited the 
Hanbury-Brown Twiss (HBT) correlations of 
hadron pairs in conjunction with fits to particle spectra, to estimate the dynamical 
properties of the reaction source in a blast wave model. The NA49 data for 
central Pb+Pb collisions at $\sqrt s_{NN} = 17.3$~GeV indicated a strong longitudinal 
flow with an approximately boost invariant longitudinal velocity profile. These data 
also suggested a transverse expansion of the pion emission source with a duration of 8-9~fm/c. 
A more recent analysis  \cite{alt07} confirms these earlier findings while extending the beam energy 
dependence of the measurements to five separate SPS energies.

Several years ago, an alternative technique based on source imaging was introduced 
for model-independent extraction of emission sources in the 
pair-center-of-mass system (PCMS) \cite{bro97,bro98,bro01}. 
This new methodology has provided a more faithful and detailed extraction of the 
actual 1D source function \cite{chu05,ppg52}. Recent theoretical 
developments  \cite{dan05,brown05,dan06,chu06} enable the extraction of 
three-dimensional (3D) profiles of the emission source.

This methodology, in both its 1D and 3D forms, has been employed for Au+Au 
reactions at $\sqrt s_{NN}=200$~GeV to obtain detailed information on 
both the spatial and the lifetime extents of the created emission 
source \cite{ppg52,ppg76}. Here, we use the new methodology to again 
study reaction dynamics at the SPS but with identified pion correlations 
and extensively developed imaging techniques that explicitly include 
Coulomb effects and do not assume Gaussian sources. The resulting 
non-Gaussian source functions are interpreted in the context of a 
powerful new simulation model, THERMINATOR \cite{therminator,kis06,kis05}. 
This approach explicitly includes all known resonance decays, longitudinal 
expansion, transverse expansion and a freeze-out hypersurface. 

 In this study, the source imaging technique is used to analyze NA49 Collaboration 
 data for central ($0-7\%$) Pb+Pb collisions 
at $\sqrt s_{NN} =$~17.3~GeV, obtained at the SPS. Model comparisons allow tests 
of different aspects of the dynamics and, in particular, the 
extraction of the proper time for breakup and emission duration for the pion 
emission source. The picture that emerges has many similarities to that from 
the early work \cite{app98,alt07}, but also adds interesting features and conclusions
that preclude direct comparison.

The data presented here were taken by the NA49 Collaboration during the years 1996-2000. 
Lead beams of 158{\it A}~GeV from the CERN SPS accelerator were made incident on a 
lead foil of thickness 224~mg/cm$^2$. Details of the experimental setup are discussed 
in Refs.\cite{alt07} and \cite{nim99}. Briefly, the NA49 Large Acceptance Hadron 
Detector achieves precision tracking and particle 
identification using four large Time Projection Chambers (TPCs). 
The first two of them are mounted in precisely mapped magnetic fields with total bending 
power of up to 9~Tm. Charged particles are detected by the tracks left in the TPCs and 
are identified by the energy deposited in the TPC gas. Mid-rapidity particle identification 
is further enhanced by a time-of-flight wall (resolution 60~ps). Charged particle momenta 
are determined from the deflection in the magnetic field. With the NA49 setup, a resolution 
of $\delta p/p^2 \approx (0.3-7)\times 10^{-4}$ (GeV/c$)^{-1}$ is achieved. 
Event centrality is determined by a forward veto calorimeter which measures the 
energy of spectator matter. Approximately 3.9 million central events were analyzed. 

The 3D correlation function, $C(\mathbf{q})$, and its 1D angle-averaged counterpart 
C($q$), were obtained as the ratio of pair to uncorrelated reference 
distributions in relative momentum $\mathbf{q}$, for $\pi^+\pi^+$ and $\pi^-\pi^-$pairs. 
Here, $\mathbf{q}=\frac{(\mathbf{p_1}-\mathbf{p_2})}{2}$ is half of the momentum difference 
between the two particles in the PCMS, and $q$ is the modulus of the vector $\mathbf q$. 
The pair distribution was obtained by pairing particles from the same event;
the uncorrelated distribution was obtained by pairing particles from different 
events. The Lorentz transformation of $\mathbf q$ from the laboratory frame to
the PCMS is made by a transformation to the longitudinally co-moving 
system (LCMS) frame along the beam direction followed by a transformation to 
the PCMS along the pair transverse momentum \cite{led04}. C($q$) is observed 
to be flat for $50 < q < 100$~MeV/c and is normalized to unity over this range.
 
Mid-rapidity ($|y_L-y_0|<$0.35, where $y_L$ and $y_0$ are particle and 
nucleus-nucleus centre-of-mass rapidities in the laboratory frame), 
low $k_T$ ($k_T<70$~MeV/c, where $k_T$ is half the transverse component 
of the pair total momentum) $\pi^+\pi^+$ and $\pi^-\pi^-$ pion pairs were 
selected for this study. Track merging and splitting effects were removed 
by appropriate cuts on both the pair and uncorrelated distributions. The merging 
cut required the two particles in a pair to be separated by at least 2.2~cm over 50 pad rows in 
the vertex TPCs \cite{alt07}. A 20$\%$ increase in this minimum separation has resulted only 
in changes within the statistical errors. Similar evaluations for the other 
cuts indicate an overall systematic uncertainty which is comparable to or smaller than the 
statistical uncertainty.
  
The effects of track momentum resolution were assessed by jittering the momenta 
of the tracks in the data by the maximum momentum resolution, 
$\delta p/p^2 \approx 7\times 10^{-4}$ (GeV/c$)^{-1}$. The resulting re-computed 1D 
and 3D correlation functions, which incorporate twice the effect of the momentum resolution, 
were found to be consistent with those obtained without momentum smearing. 
The correlation functions without additional smearing serve in the following as a basis for the extraction of source functions 
via imaging and fitting.   


The imaging procedure employed uses the 1D imaging code of Brown and 
Danielewicz \cite{bro97,bro98,bro01}, which has been successfully used 
to image 1D correlation functions obtained at 
$\sqrt s_{NN}=200$~GeV \cite{ppg52}. Briefly, the technique numerically inverts 
the 1D Koonin-Pratt equation,
\begin{equation}
  C(q)-1=R(q)=4\pi\int dr r^2 K_0(q,r) S(r)
  \label{kpeqn}
\end{equation}
which relates the two-particle angle-averaged 1D correlation 
function, C($q$), to the 1D source function or image, S($r$).
The latter gives the probability of emitting a pair of 
particles with a separation distance $r$ in the PCMS. The 1D kernel $K_0(q,r)$ 
incorporates the effects of Coulomb interaction and of Bose-Einstein symmetrization.

Contamination by uncorrelated pairs (weak decay products accepted by the track 
selection cuts, misidentified particles, etc.) dilute the correlation and 
reduce $R(q)$. It has been confirmed by simulation that the contamination is 
approximately constant in $q$, so that the reduction factor can be assumed 
to be $q$-independent. The source function $S(r)$ then gets reduced by the 
same r-independent factor due to the linearity of Eq.~(\ref{kpeqn}).

\begin{figure}
\includegraphics[width=1.0\linewidth]{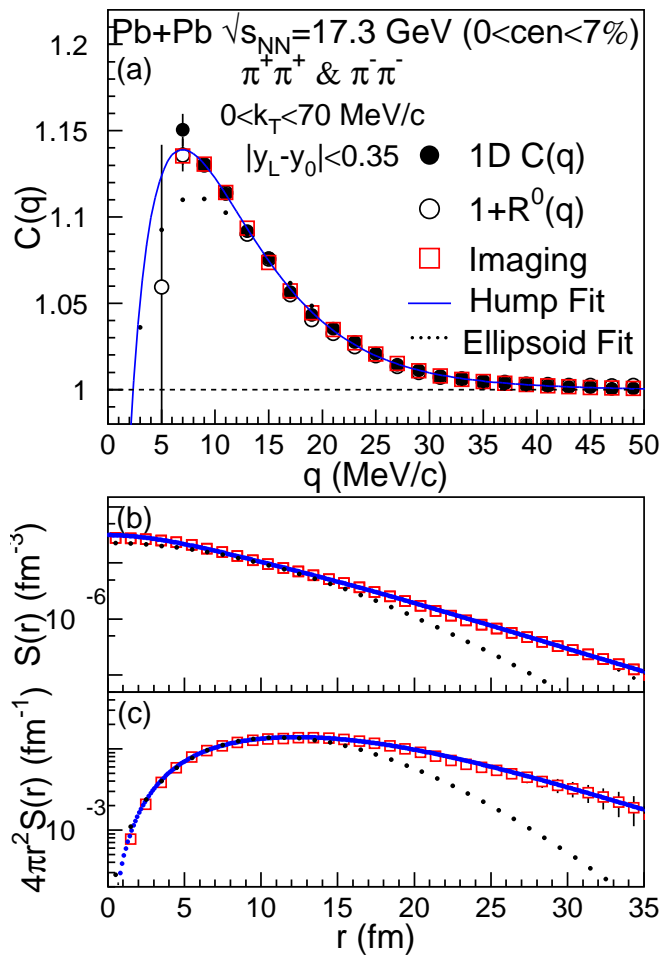}
 \caption{\label{na49_fig1_ppg}
{(color on line) Angle-averaged correlation function (top panel), source function 
(middle) and radial probability density (bottom) for mid-rapidity pion pairs. Filled circles 
show correlation from direct averaging of the data. Error bars indicate statistical errors only; systematic uncertainties are smaller than statistical ones. Open circles represent correlation
from fitting the data using angular decomposition. Squares show the imaged source and 
correlation corresponding to the imaged source. The dotted and solid lines represent, respectively, the fitted Gaussian and Hump Eq.~(\ref{ss_eqn}) sources and their corresponding correlation functions.}
}
\end{figure}

Figure~\ref{na49_fig1_ppg}(a) shows data points for the 1D correlation function in 
relation to the imaged source function in Figs.~\ref{na49_fig1_ppg}(b) and (c), for mid-rapidity, low $p_T$ pion pairs.
The source function indicates a tail for $r\agt 15$~fm which is qualitatively 
similar to that reported for RHIC data in Ref.\cite{ppg52}. 
As a check, the extracted source function is used as input to Eq.~(\ref{kpeqn}) 
to obtain a restored correlation function also shown in Fig.~\ref{na49_fig1_ppg}(a);  
excellent consistency is observed. 

	In parallel to the imaging procedure, two different functional forms 
were used to fit the measured correlation function directly, as discussed below. 
The conclusion from the fits (see Fig.~\ref{na49_fig1_ppg}) is that 
a triaxial Gaussian, frequently termed ellipsoid, as used in traditional  
HBT methodology, poorly describes the correlation function at low 
$q\alt 13$~MeV/c (Fig.~\ref{na49_fig1_ppg}(a)), 
and this leads to a deviation from the tail of the imaged source function 
for large $r\agt 15$~fm (Fig.~\ref{na49_fig1_ppg}(b)). Fig.~\ref{na49_fig1_ppg}(c) 
highlights the fact that the tail for $r\agt 15$~fm contains a very significant 
fraction of the source.
On the other hand, the Hump function (cf. Eq.~(\ref{ss_eqn}) and discussion below) gives a good 
fit over a more extensive range.

For systematic access to the 3D source function $S(\mathbf{r})$, the 3D correlation 
function $C(\mathbf{q})$ and source function $S(\mathbf{r})$ were both expanded in
a series with correlation moments 
$R^l_{\alpha_1 \ldots \alpha_l}(q)$ and source moments $S^l_{\alpha_1 \ldots \alpha_l}(r)$ 
 in a Cartesian harmonic basis representation:
\begin{equation}
C(\mathbf{q}) - 1 = R(\mathbf{q}) = \sum_l \sum_{\alpha_1 \ldots \alpha_l}
   R^l_{\alpha_1 \ldots \alpha_l}(q) \,A^l_{\alpha_1 \ldots \alpha_l} (\Omega_\mathbf{q}),
 \label{eqn1}
\end{equation}

\begin{equation}
 S(\mathbf{r}) = \sum_l \sum_{\alpha_1 \ldots \alpha_l}
   S^l_{\alpha_1 \ldots \alpha_l}(r) \,A^l_{\alpha_1 \ldots \alpha_l} (\Omega_\mathbf{r}),
\label{eqn3}
\end{equation}
where $l=0,1,2,\ldots$, $\alpha_i=x, y \mbox{ or } z$, $A^l_{\alpha_1 \ldots \alpha_l}(\Omega_\mathbf{q})$
are Cartesian harmonic basis elements ($\Omega_\mathbf{q}$ is the solid angle in 
$\mathbf{q}$ space) and $R^l_{\alpha_1 \ldots \alpha_l}(q)$ are Cartesian correlation moments given by
\begin{equation}
 R^l_{\alpha_1 \ldots \alpha_l}(q) = \frac{(2l+1)!!}{l!}
 \int \frac{d \Omega_\mathbf{q}}{4\pi} A^l_{\alpha_1 \ldots \alpha_l} (\Omega_\mathbf{q}) \, R(\mathbf{q})
 \label{eqn2}
\end{equation}
Here, the coordinate axes are oriented so that $z$ (long) is parallel to the beam direction, 
$x$  (out) points in the direction of the total momentum of the pair in the LCMS frame 
and $y$ (side) is chosen to form a right-handed coordinate system with $x$ and $z$.
   
\begin{figure}
\includegraphics[width=1.0\linewidth]{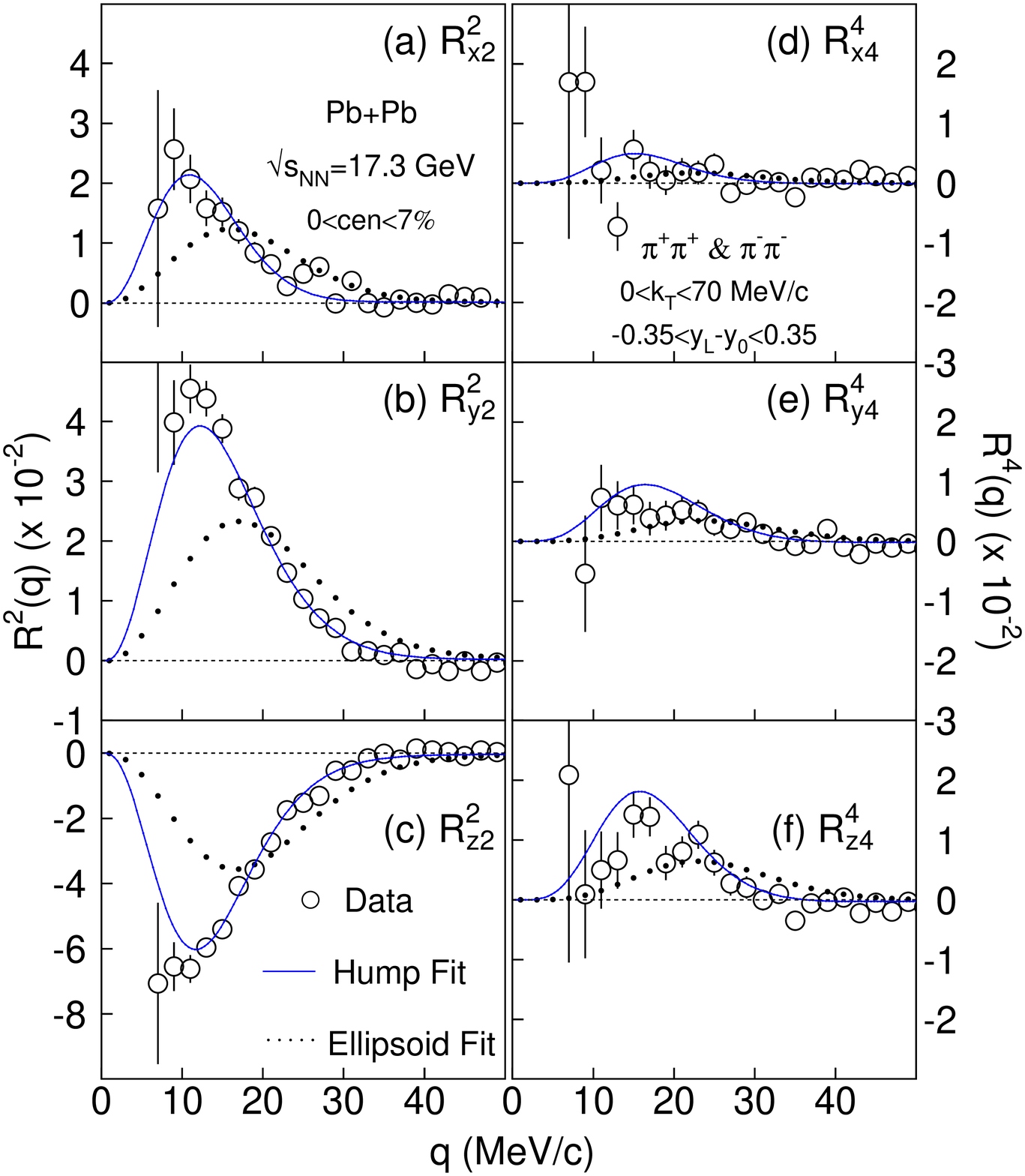}
 \caption{\label{na49_fig2_ppg}
{(color online) Correlation moments for multipolarity $l=2$ (left panels), 
and $l=4$ (right panels) for mid-rapidity $\pi^+\pi^+$ and $\pi^-\pi^-$ pairs.
Error bars indicate statistical errors only; systematic uncertainties are smaller than statistical ones.}
}
\end{figure}

The correlation moments, for each order $l$, can be calculated from the measured 3D correlation function 
using Eq.~(\ref{eqn2}). Alternatively, Eq.~(\ref{eqn1}) can be truncated so as to include all non-vanishing 
moments and expressed in terms of independent moments only. As expected from symmetry considerations, moments odd in any coordinate were found to be consistent with zero within
statistical uncertainty. Up to order $l=4$, there are 6 independent 
moments: $R^0$, $R^2_{x2}$, $R^2_{y2}$, $R^4_{x4}$, $R^4_{y4}$ and $R^4_{x2y2}$, 
where $R^2_{x2}$ is shorthand for $R^2_{xx}$ etc. The independent moments
can then be extracted as a function of $q$ by fitting the truncated series to the
experimental 3D correlation function with the moments as the parameters of the fit. 
The present analysis emphasizes the second method, with the moments computed up to 
order $l=4$ (higher order moments are found to be negligible).
The moments are shown in Fig.~\ref{na49_fig2_ppg}, for the multipolarity orders of $l= $2 and 4,
and in Fig. \ref{na49_fig1_ppg}a for $l=0$ ($1+R^0(q)\equiv C^0(q)$) \cite{compare_cq_c0q}. The magnitude of the moment $R^4_{x2y2}$ is comparable to that of $R^4_{x4}$ and $R^4_{y4}$. 

Substitution of $C(\mathbf q)$ and $S(\mathbf r)$, from Eq.~(\ref{eqn1}) 
and Eq.~(\ref{eqn3}), into the 3D form of the Koonin-Pratt equation
\begin{equation}
  C(\mathbf{q})-1 = \int d\mathbf{r} K(\mathbf{q},\mathbf{r}) S(\mathbf{r})
  \label{3dkpeqn}
\end{equation}
results \cite{dan05} in a relationship between corresponding correlation $R^l_{\alpha_1 \ldots \alpha_l}(q)$ and source moments $S^l_{\alpha_1 \ldots \alpha_l}(r)$,
which is similar to the 1D Koonin-Pratt equation:
  
\begin{equation}
  R^l_{\alpha_1 \ldots \alpha_l}(q) = 4\pi\int dr r^2 K_l(q,r) S^l_{\alpha_1 \ldots \alpha_l}(r),
  \label{momkpeqn}
\end{equation}
but now pertains to moments describing different ranks of angular anisotropy $l$. 
Since the mathematical structure of Eq.~(\ref{momkpeqn}) is the same as that 
of Eq.~(\ref{kpeqn}), the same imaging technique can be used to invert the kernel $K_l$ of the relation to extract the source
moment $S^l_{\alpha_1 \ldots \alpha_l}(r)$ from the corresponding correlation moment $R^l_{\alpha_1 \ldots \alpha_l}(q)$. 
Finally, the total 3D source function is calculated by combining the source moments for 
each $l$ as in Eq.~(\ref{eqn3}).

Alternatively, the source function can be extracted by directly fitting the 
3D correlation function with an assumed 3D shape for the source function. Since 
the 3D correlation function can be represented by the Cartesian moments in the 
harmonic decomposition, the 3D fit corresponds to fitting the six independent non-trivial moments 
simultaneously with a trial source function. 

	Figures~\ref{na49_fig1_ppg}-\ref{na49_fig2_ppg}
show the result of direct fits to the independent correlation moments with two 
3D functions: (a) a single triaxial Gaussian, or ellipsoid, (dotted curve) and (b) a Hump 
shape (solid curve). As mentioned, the ellipsoidal fit, with four free parameters, fails to capture 
the low $q$ behavior in C($q$) and the large $r$ behavior in S($r$). 
On the other hand, the Hump function, with six free parameters, gives a good fit. The form of the Hump function is
\begin{equation}
  S(x,y,z) = \Lambda \exp\left[-f_s \frac{r^2}{4 r_{s}^2} - f_l\left(\frac{x^2}{4 r_{xl}^2}  
  + \frac{y^2}{4 r_{yl}^2} + \frac{z^2}{4 r_{zl}^2}\right)\right]
  \label{ss_eqn}
\end{equation}
where $r^2=x^2+y^2+z^2$ and the coefficients $f_s$ and $f_l$ of the short and 
long-ranged components are given by $f_s=1/[1+(r/r_0)^2)]$ and $f_l = 1-f_s$ 
respectively. Here, the argument of the exponential shifts the behavior from 
that of a simple spherically symmetric Gaussian for $r\ll r_0$ to that of a 
triaxial Gaussian for $r\gg r_0$. The parameter $\Lambda$ regulates the fraction 
of pion pairs of which correlations are described in terms of the Hump 
function (for fit parameter values see Ref.~\cite{hump_function}).

Source imaging involves no assumptions on the analytical shape of 
the 3D source function. On the other hand, the moment fitting explicitly invokes a particular form 
for the 3D source function. The ellipsoid fit produces a $\chi^2/ndf$ value of 6.8 
while the Hump produces 1.2, which indicates a better fit to the observed 
correlation moments, as is visually evident in Figs. \ref{na49_fig1_ppg}(a) and \ref{na49_fig2_ppg}. 
Close agreement between 
the experimental data, the Hump fit and the restored correlation moments from imaging 
(see Figs.~\ref{na49_fig1_ppg}-\ref{na49_fig2_ppg}) strongly suggests that this assumed 
functional form properly represents the emission source. However, the uniqueness of the
source function is, for example, not guaranteed beyond the region to which data are sensitive such as $r>40$~fm or where the source function is very small.

\begin{figure}
\includegraphics[width=1.0\linewidth]{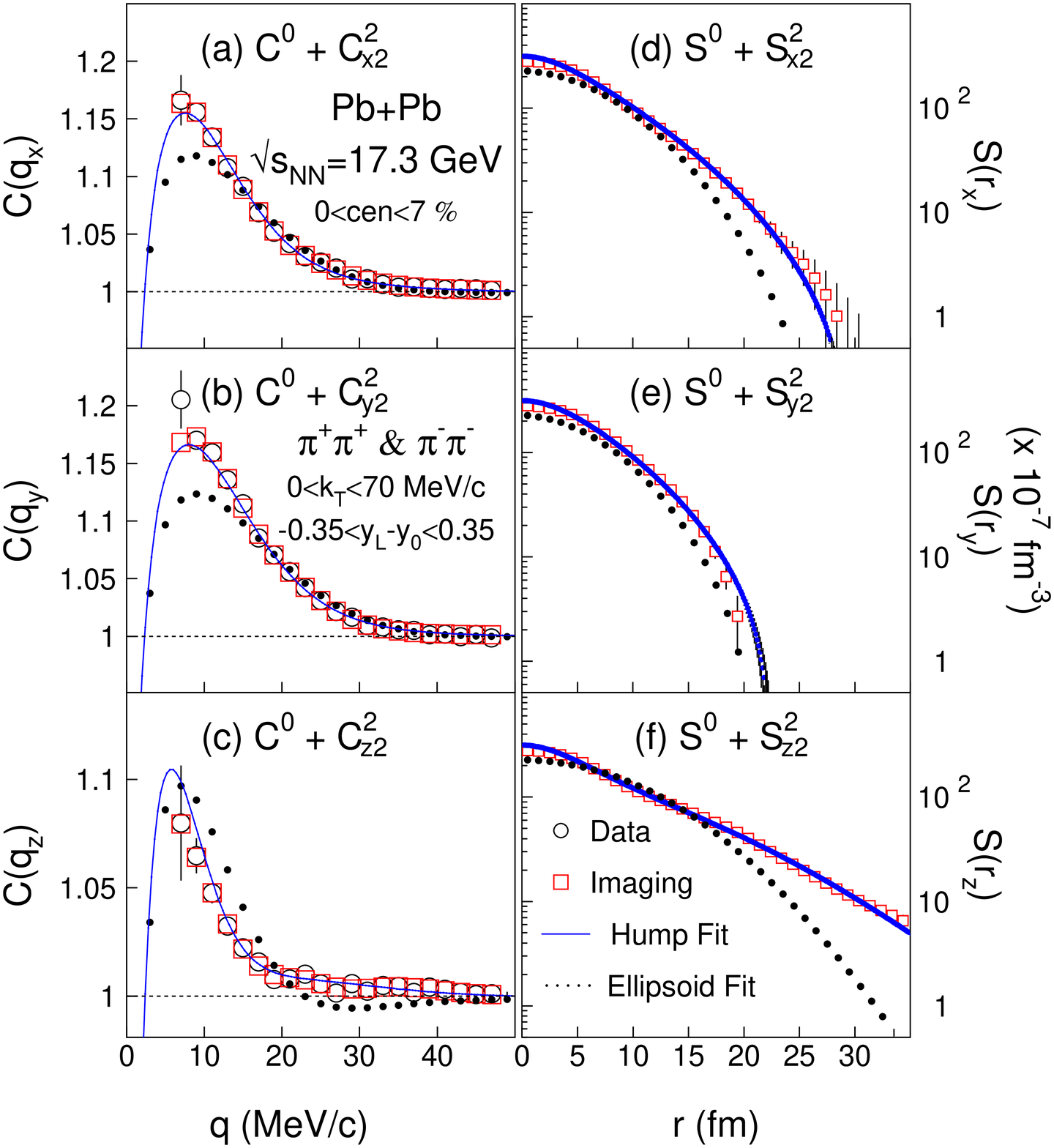}
 \caption{\label{na49_fig3_ppg}
{(color online) Correlation $C(q_i)$ (left panels) and source $S(r_i)$ (right panels)
function profiles for $\pi^+\pi^+$ and $\pi^-\pi^-$ pairs in the outward $x$ (top panels), sideward $y$
(middle) and longitudinal $z$ (bottom) directions. The use of symbols is analogous to that 
in Fig.~\ref{na49_fig1_ppg}. Error bars indicate statistical errors only; systematic uncertainties are 
smaller than statistical ones. Here, $l=4$ moments make negligible contributions.}
}
\end{figure}

Figures~\ref{na49_fig3_ppg}(d)-(f) show comparisons between two-pion source functions 
obtained via the fitting (lines) and the imaging (squares) techniques. The ellipsoid 
fit function (dotted line) underestimates the source image (squares) and Hump 
fit function (solid line) for $r>15$~fm in the $x$ and $z$ directions while the 
Hump fit function is in good agreement with the source image in 
the $x$, $y$ and $z$ directions. This consistency check emphasizes the high degree 
of integrity with which the 3D source function is being extracted. The source 
function in the $z$ direction is characterized by a long tail which extends beyond 30~fm. 
The source function in $x$ also has a non-Gaussian tail, which, for 
this low $p_T$ cut is less prominent than that in $z$. 
These aspects are decidedly different from those of a RHIC study \cite{sqm07}.

The difference between the source functions from the ellipsoid fit and imaging 
procedures is also evident from a comparison of the corresponding correlation functions 
in the $x$, $y$ and $z$ directions as shown in Fig.~\ref{na49_fig3_ppg}(a)-(c) respectively. 
Again there is consistency between the data, Hump fit and restored correlation 
functions in all three directions while the differences between the ellipsoid and 
Hump fit sources for $r \agt 15$~fm are manifest by differences between 
the respective correlation functions for $q\alt 15$~MeV/c.

	The event simulation code THERMINATOR allows for tests of the emission dynamics and of
the breakup time of the reaction systems \cite{chu07,therminator,kis06,kis05}. 
The code simulates thermal emissions from a cylinder with input 
transverse radius $\rho_{\text{max}}$. Bjorken longitudinal 
boost invariance is assumed, and an expansion with transverse
radial velocity $v_r(\rho)=(\rho/\rho_{\text{max}})/(\rho/\rho_{\text{max}}+ v_t)$, 
where $v_t=1.41$, in the Blast-Wave mode of the code. 
A fluid element ring, defined by $\rho$ and $z$, breaks up
at proper time $\tau$ and lab frame time $t$ where $t^2 = \tau^2 + z^2$. The freezeout
hypersurface is specified by $\tau = \tau_0 + a\rho$ where $a$, the space-time 
correlation parameter, is set to -0.5 as was found in Ref. \cite{kis05}. 
The negative value of $a$ implies ``outside-in'' burning of the source 
i.e outer particles are emitted earlier than inner ones, while a positive 
value of $a$ would imply the reverse i.e source emission from inside out. 
An emission duration parameter $\Delta\tau$ is also needed to achieve a good 
fit. All known hadronic resonance decays are included.

THERMINATOR parameters $v_t,\; T,\; \mu_B,\; \mu_s,\; \mu_i$ and $a$ are taken 
from Ref.\cite{therminator,kis06,kis05,bro02} as obtained from spectra and particle 
yields. Values of $\rho_{max},\; \tau_0$ and $\Delta \tau$ were obtained by matching
THERMINATOR's generated source function to data shown in Figs.~\ref{na49_fig3_ppg}(d - f).
The value of the transverse radius $\rho_{max}$ is chosen so as to reproduce 
the source function profile in the $y$ direction; S($r_y$) is insensitive to $\tau_0$ 
and $\Delta \tau$. The proper lifetime $\tau_0$ is determined by the short-range 
behavior of the source function profiles in the $x$ and $z$ directions. 
The proper emission duration is then determined by the tails of the source 
profiles in the $x$ and $z$ directions.  

\begin{figure}
\includegraphics[width=1.0\linewidth]{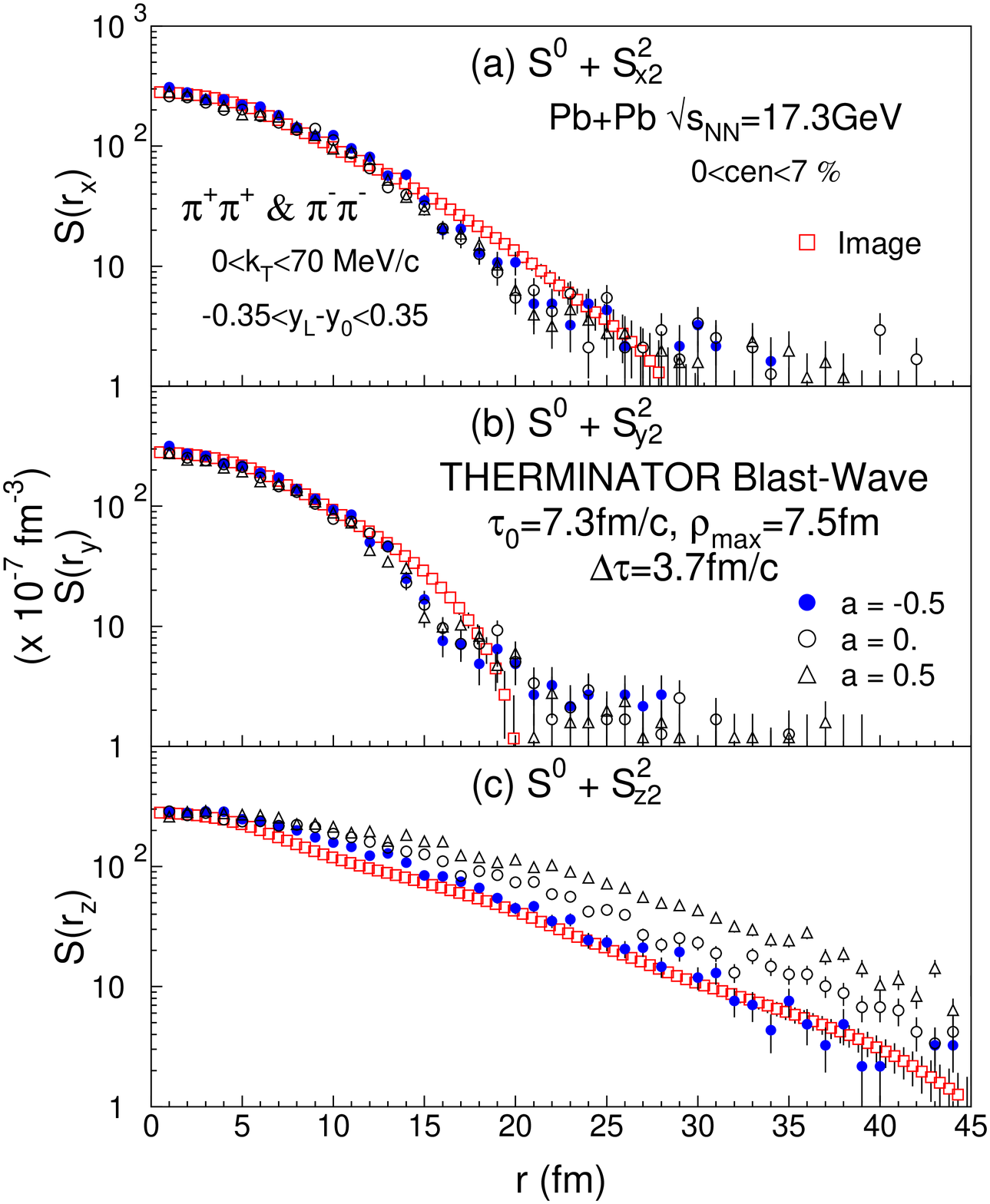}
 \caption{\label{na49_fig4_ppg}
{(color online) Source function profiles, S($r_i$), comparison in the 
(a) $x$, (b) $y$ and (c) $z$ directions between the imaged data (squares) 
and THERMINATOR Blast-Wave model with various values of $a$ 
(circles and triangles). Error bars indicate statistical errors only; 
systematic uncertainties are smaller than statistical ones.}
}
\end{figure}

The calculation gives a good match to the experimental source function in 
the $x$, $y$ and $z$ directions with a transverse dimension $\rho_{max}=7.5\pm0.1$~fm, 
proper lifetime $\tau_0$ ($\tau=\tau_0$ at $\rho=0$) of $7.3\pm0.1$ fm/c, a proper 
emission duration $\Delta\tau=3.7\pm0.1$~fm/c and $a=-0.5$ (solid circles) \cite{volume}. 
The errors quoted are from the matching procedure alone. 
With these values of $\rho_{max},\; \tau_0$ and $\Delta \tau$ we have reexamined 
the role of $a=-0.5$ i.e. outside-in burning. Figure~\ref{na49_fig4_ppg} shows a 
comparison of the THERMINATOR source function, calculated using various values 
of $a$ and other previously tuned parameters \cite{bro02}, with the extracted 
source function. The open symbols show that the calculations with $a\ge0$ overstate 
the extracted source function profile in the $z$ direction. 
Attempts to compensate for this overshoot via different combinations of $\rho_{max}$,
$\tau_0$ and $\Delta\tau$ were unsuccessful. Therefore, this failure 
suggests that a negative value for $a$, hence ``outside-in'' particle emission, 
is required to reproduce the extracted source function.  The success of the THERMINATOR 
model simulation in precisely reproducing the experimental source function 
indicates consistency with approximate boost invariance at mid-rapidity, blast-wave dynamics 
for transverse flow, and outside-in burning in the evolution of the expanding 
system.  

Results from this study and those from Ref.\cite{alt07} depend on the 
different analysis techniques and models employed. The deduced time 
scales are similar but the geometric transverse radius is quite different. 
This difference results from the inclusion of resonances in THERMINATOR, as well 
as different parametrizations of $T$ and $v_r(\rho)$. Conclusions from these THERMINATOR
parameters are, of course, model dependent and therefore not necessarily unique. 
Different model assumptions may possibly lead to different pictures of the reaction 
dynamics \cite{Lisa:2008gf}.  

In summary, we have presented a three-dimensional femtoscopic study of the two-pion 
source function in Pb+Pb collisions at $\sqrt s_{NN}$ = 17.3~GeV. A model-independent 
imaging/fitting technique reveals prominent non-Gaussian tails in the outward and 
longitudinal directions of the extracted source function. THERMINATOR Blast-Wave model 
calculations, incorporating Bjorken longitudinal flow, give a near-exponential 
tail in the longitudinal direction consistent with observation. 
The space-time correlation parametrization suggests outside-in burning and provides 
values of the proper time for breakup and the emission duration.


 This work was supported by the US Department of Energy
Grant DE-FG03-97ER41020/A000 and DE-FG02-03ER41259,
the Bundesministerium fur Bildung und Forschung, Germany 06F 137,
the Virtual Institute VI-146 of Helmholtz Gemeinschaft, Germany,
the Polish State Committee for Scientific Research (1 P03B 006 30, 1 P03B 097 29, 1 PO3B 121 29, 1 P03B 127 30),
the Hungarian Scientific Research Foundation (T032648, T032293, T043514),
the Hungarian National Science Foundation, OTKA, (F034707),
the Polish-German Foundation, the Korea Science \& Engineering Foundation (R01-2005-000-10334-0),
the Bulgarian National Science Fund (Ph-09/05), the Croatian Ministry of Science, Education and Sport (Project 098-0982887-2878), the National Science Foundation NSF PHY-0555893 and PHY-0800026.


\begin{thebibliography}{0}
\expandafter\ifx\csname natexlab\endcsname\relax\def\natexlab#1{#1}\fi
\expandafter\ifx\csname bibnamefont\endcsname\relax
  \def\bibnamefont#1{#1}\fi
\expandafter\ifx\csname bibfnamefont\endcsname\relax
  \def\bibfnamefont#1{#1}\fi
\expandafter\ifx\csname citenamefont\endcsname\relax
  \def\citenamefont#1{#1}\fi
\expandafter\ifx\csname url\endcsname\relax
  \def\url#1{\texttt{#1}}\fi
\expandafter\ifx\csname urlprefix\endcsname\relax\def\urlprefix{URL }\fi
\providecommand{\bibinfo}[2]{#2}
\providecommand{\eprint}[2][]{\url{#2}}

\end{thebibliography}


\begin{thebibliography}{10}


\bibitem{shu05} E.V.~Shuryak, Nucl. Phys. {\bf A 750}, 64 (2005).
\bibitem{pra84} S.~Pratt, Phys. Rev. Lett. {\bf 53}, 1219 (1984); K.~Kolehmainen and M.~Gyulassy, Nucl. Phys. {\bf A 461}, 239c (1987), Phys. Lett. {\bf B 180} 203 (1986); A.~Makhlin and Y.~Sinyukov, Z. Phys. {\bf C 39} 69 (1988).
\bibitem{adc05} K.~Adcox et al., Nucl. Phys. {\bf A757}, 184 (2005).
\bibitem{lis05} G.~Goldhaber et al., Phys. Rev. 120 (1960) 300; G.~I.~Kopylov and M.~I.~Podgoretsky, Sov.J.Nucl.Phys. {\bf 15}, 219 (1972); G.~Cocconi, Phys. Lett. {\bf 49 B} (1974) 459; M.~Lisa et al., Annu.Rev.Nucl.Part.Sci.{\bf 55}, 357 (2005).
\bibitem{Chapman:1996ec}
  S.~Chapman and J.~R.~Nix,
  Phys.\ Rev.\  C {\bf 54}, 866 (1996)
\bibitem{app98} H.~Appelsh\"aeuser et al., Eur.Phys.J.,{\bf C2}, 661 (1998).
\bibitem{alt07} C.~Alt et al., Phys. Rev. {\bf C 77}, 064908 (2008).
\bibitem{bro97} D.~Brown, P.~Danielewicz, Phys. Lett. {\bf B 398}, 252 (1997).
\bibitem{bro98} D.~Brown, P.~Danielewicz, Phys. Rev. {\bf C 57}, 2474 (1998).
\bibitem{bro01} D.~Brown, P.~Danielewicz, Phys. Rev.{\bf C 64}, 14902 (2001).
\bibitem{chu05} P.~Chung et al., Nucl. Phys. {\bf A 749}, 275c (2005).
\bibitem{ppg52} S.~Adler et al., Phys. Rev. Lett. {\bf 98}, 132301 (2007).
\bibitem{dan05} P.~Danielewicz and S.~Pratt, Phys.Lett.{\bf B 618} 60 (2005).
\bibitem{brown05} D.A.~Brown et al., Phys. Rev. {\bf C 72}, 054902 (2005).
\bibitem{dan06} P.~Danielewicz and S.~Pratt, nucl-th/0612076v2 (2007).
\bibitem{chu06} P.~Chung et al., Nucl. Phys. {\bf A 774}, 919 (2005).
\bibitem{ppg76} S.~Afanasiev et al., Phys. Rev. Lett. {\bf 100}, 232301 (2008).
\bibitem{therminator} A.~Kisiel et al., Comput. Phys. Commun. 174, 669, 2006.
\bibitem{kis06} A.~Kisiel, Brazilian Journal Physics, {\bf37}, 3A, 917 (2007).
\bibitem{kis05} A.~Kisiel et al., Phys. Rev. {\bf C 73}, 064902 (2006).
\bibitem{nim99} S.~Afanasiev et al., Nucl. Inst. Methods {\bf A} 430, 210 (1999).
\bibitem{led04} R.~Lednicky et al., Phys. Part. Nucl. {\bf 35}, S50 (2004).
\bibitem{compare_cq_c0q} $C^0(q)$ agrees with 1D correlation function $C(q)$ attesting 
         to the reliability of the moment extraction technique. 
\bibitem{hump_function} The fit parameters for the Hump are $\Lambda = 0.281\pm 0.006$
         $r_0=5.8\pm0.3,\; r_s=2.5\pm0.1,\; r_{xl}=6.9\pm0.1,\; r_{yl}=6.0\pm0.1,\; r_{zl}=10.9\pm 0.3$.
         Those for the ellipsoid (i.e. $f_s \equiv r_0 =0$) are $\Lambda =0.198\pm0.003, r_{xl}=5.46\pm0.04,
         \; r_{yl}=4.95\pm0.04,\; r_{zl}=7.67\pm0.08$.
\bibitem{sqm07} P.~Chung et al, J.~Phys.~G:~Nucl.~Part.~Phys.~{\bf 34} 1 (2007)
\bibitem{chu07} P.~Chung and P.~Danielewicz, arXiv:0807.4892v1 [nucl-ex] 30 Jul 2008.
\bibitem{bro02} W.~Broniowski and W.~Florkowski, arXiv:hep-ph/0202059v1 7 Feb 2002; M.~Michalec, 
Ph.D. Thesis, nucl-th/0112044; Values used are: $T=164$~MeV for temperature, $\mu_B=229$~MeV, 
$\mu_S=54$~MeV, $\mu_I=-7$~MeV for baryon, strangeness and isospin chemical potentials.
\bibitem{volume} For this parameter set, THERMINATOR produces $m_T$ spectra which are in very good 
                 quantitative agreement with the measurements reported 
                 in Phys. Rev.{\bf C 66}, 054902 (2002). 


\bibitem{Lisa:2008gf}
  M.~A.~Lisa and S.~Pratt, arXiv:0811.1352 [nucl-ex].


\end{thebibliography}
\end{document}